\documentclass[twocolumn]{aastex63}

\usepackage{amsmath}
\usepackage{wasysym}
\usepackage{graphicx}  
\usepackage{hyperref}
\usepackage{natbib}
\usepackage{xcolor}
\usepackage{url}
\RequirePackage{lineno}

\def\teff{{T$_{\text{eff}}$\,\,}}

\def\phos{PO$_4^{3-}$}

\def\gtaprx{ \mathrel{ \vcenter{
      \offinterlineskip \hbox{$>$}
      \kern 0.3ex \hbox{$\sim$}    } } }

\def\ltaprx{ \mathrel{ \vcenter{
      \offinterlineskip \hbox{$<$}
      \kern 0.3ex \hbox{$\sim$}    } } }
      
\def\aj{{AJ}}
\def\apj{{ApJ}}
\def\apjs{{ApJS}}
\def\apjl{{ApJL}}

\def\aap{{A\&A}}

\shorttitle{Phosphorus in Stars, Planets, and Astrobiology}
\shortauthors{Hinkel et al.}

\begin{document}

\title{The Influence of Stellar Phosphorus On Our Understanding of Exoplanets and Astrobiology}

\author[0000-0003-0595-5132]{Natalie R.\ Hinkel}
\affiliation{Southwest Research Institute, 6220 Culebra Rd, San Antonio, TX 78238, USA}

\author[0000-0003-0736-7844]{Hilairy E. Hartnett}
\affiliation{School of Earth and Space Exploration, Arizona State University, Tempe, AZ 85287, USA}
\affiliation{School of Molecular Sciences, Arizona State University, PO Box 871604, Tempe, AZ 85287}

\author[0000-0003-1705-5991]{Patrick A. Young}
\affiliation{School of Earth and Space Exploration, Arizona State University, Tempe, AZ 85287, USA}

\correspondingauthor{Natalie Hinkel \& Hilairy Hartnett}
\email{natalie.hinkel@gmail.com, h.hartnett@asu.edu}

\begin{abstract}

When searching for exoplanets and ultimately considering their habitability, it is necessary to consider the planet's composition, geophysical processes, and geochemical cycles in order to constrain the bioessential elements available to life. Determining the elemental ratios for exoplanetary ecosystems is not yet possible, but we generally assume that planets have compositions similar to those of their host stars. Therefore, using the Hypatia Catalog of high-resolution stellar abundances for nearby stars, we compare the C, N, Si, and P abundance ratios of main sequence stars with those in average marine plankton, Earth's crust, as well as bulk silicate Earth and Mars. We find that, in general, plankton, Earth, and Mars are N-poor and P-rich compared with nearby stars. However, the dearth of P abundance data, which exists for only $\sim$1\% of all stars and 1\% of exoplanet hosts, makes it difficult to deduce clear trends in the stellar data, let alone the role of P in the evolution of an exoplanet. Our Sun has relatively high P and Earth biology requires a small, but finite, amount of P. On rocky planets that form around host stars with substantially less P, the strong partitioning of P into the core could rule out the potential for surface P and, consequently, for life on that planet's surface. Therefore, we urge the stellar abundance community to make P observations a priority in future studies and telescope designs. 

\end{abstract}


\section{Introduction \& Background}
\subsection{Exoplanets and life}
Planets are now known to be commonplace. To date, there are thousands of confirmed planets around other stars and perhaps as many if not more candidate planets yet to be confirmed\footnote{NASA Exoplanet Archive \url{https://exoplanetarchive.ipac.caltech.edu/}}. The very existence of planets beyond our Solar System compels us to ask whether there might be life elsewhere in the galaxy. The search for life on exoplanets is fundamentally a chemical search for atmospheric gases produced by life \citep{Seager13, DesMarais02}; for the foreseeable future this will continue to be the case \citep{DomagalGoldman16, Schwieterman18}. 

On Earth, life both responds to its chemical environment and imparts a chemical signature on its environment \citep{Shock15}. Because we understand some of the chemical conditions conducive to life on our planet, we presume that we can infer something about life on other planets from their observed compositions. Chemical compositions are inherently linked to the geochemical and geophysical processes that are, in turn, connected to the stellar elemental abundances, creating highly interdisciplinary studies.

\subsection{Phosphorus is a key element for life}
On Earth, the key elements for biology are H, C, N, O, P, S, (or CHNOPS) as well as a few alkali and alkaline earth metals, and a handful of transition metals. In particular phosphorus (as orthphosphate, \phos) is necessary for all life. It forms the structural backbone of genetic molecules \citep[i.e., DNA and RNA; ][]{schlesinger13} and it is the energy currency of nearly all metabolism \citep[i.e., ATP;][]{Lehninger17}. In modern marine settings, phosphorus is considered to be the ultimate limiting nutrient for life; i.e., it is the chemical species least available relative to the molar requirements for biochemical reactions. Phosphorus is limiting as opposed to nitrogen or carbon (also key elements for biology) because the only source of P is rock weathering; in contrast, microbes have evolved the ability to fix biologically useful forms of nitrogen and carbon from the gas phase \citep{schlesinger13}. The biological need for C, N, and P in relatively fixed molar ratios has been known for decades \citep{Redfield1958, redfieldetal63}. The seminal \citet{redfield34} paper was the first to report the C:N:P ratio in marine plankton and demonstrate that it is the same as the C:N:P ratio of dissolved ions in the ocean. The Redfield ratio of 106 carbons for every 16 nitrogen and every 1 phosphorous (i.e., 106:16:1::C:N:P) is a remarkably robust relationship for marine plankton. This proportion summarizes the chemical stoichiometry of the simplified equation for oxygenic photosynthesis (Eq. \ref{eq:photo}) where plants use CO$_2$, nutrients, and water (left side of the equation) to produce organic matter and oxygen gas (right side of the equation). In this case, the CHNOPS species presented is a representation of biomass. This fundamental chemical relationship links the chemical composition of environments and life's processe and become the theoretical basis for the field of ecological stoichiometry \citep{sterner02}. The concept has also been expanded beyond just C, N, and P to include other elements such as iron, silicon, potassium, etc. \citep[e.g.,][]{Hoetal03}. 

\begin{equation}
\begin{split}
    106 \mathrm{CO}_2 \, + \, &16\mathrm{NO}_{3-} \, + \, 1 \mathrm{HPO}_4^{2-} \, + \, 122 \mathrm{H}_2\mathrm{O} \, + \, 18\mathrm{H}^+  \\
    &\rightarrow \mathrm{C}_{106}\mathrm{H}_{263}\mathrm{O}_{111}\mathrm{N}_{16}\mathrm{P}_1 \, + \, 138 \, \mathrm{O}_2
\end{split}\label{eq:photo}
\end{equation}

\subsection{Exoplanetary surfaces}
It is not currently possible to measure the surface composition of an exoplanet. Direct compositional measurements of the planet are limited to atmospheric spectroscopy which can only occur during a transit. 
Yet, knowing the composition of the planet's atmosphere is only one factor in determining whether a planet is habitable from a chemical perspective. It is absolutely vital that an exoplanet have surface water as well as exposed continental rock, which would ensure important geochemical cycles such as subaerial weathering, that can make available and replenish important bioessential elements necessary for life \citep{Glaser20}.

Therefore, until such time as we are able to measure the surface composition of an exoplanet (i.e., via direct imaging reflectance/emission spectroscopy or more classical ``ground truthing"), we must utilize the composition of the star as a proxy for the planet's make-up. Since stars and planets are formed at the same time within the stellar birthcloud, it is reasonable to assume a 1:1 correlation between abundance of elements in the star and the material out of which planets form \citep{Thiabaud15}. Of course, there are some exceptions to this relationship in final planet compositions that depend on the element (volatile vs. refractory) and distance the planet forms from the host star (i.e., before or beyond the ice line). For example, the Sun, Earth, and Mars all agree to within 10\% in the relative proportions of the major rocky planet building elements. However, the chemical connection between star and planet offers a starting point for modeling the interior composition and mineralogy of an exoplanet \citep{bond_2010_aa, Hinkel18}.

\section{Stellar Elemental Abundance Data}
The Hypatia Catalog is the largest database of high resolution stellar abundances for nearby stars \citep{Hinkel14} and is composed of +350,000 abundance measurements\footnote{All abundances can be accessed at \url{www.hypatiacatalog.com}}. 
The Hypatia Catalog contains +77 elements within $\sim$9400 main sequence (FGKM-type) stars, all of which are within 500 pc (1.03$\times$10$^8$ AU) of the Sun; all exoplanet host stars are included regardless of distance. As part of its latest update, all stars within the Hypatia Catalog are cross-matched to SIMBAD, the NASA Exoplanet Archive, Gaia, and the TESS Input Catalog for the most up-to-date stellar properties. 
Stellar abundances are usually reported such that each element, A, is normalized to $10^{12}$ H atoms: log $\epsilon$(A) = log(N$_A$/N$_H$)+12. Then, the element ratio in the star is normalized with respect to the same ratio in the Sun and are indicated by square brackets: [A/B] = log(N$_A$/N$_B$)$_*$ - log(N$_A$/N$_B$)$_{\odot}$ dex. 

\begin{figure}[h]
    \centering
    \includegraphics[width=1.0\linewidth]{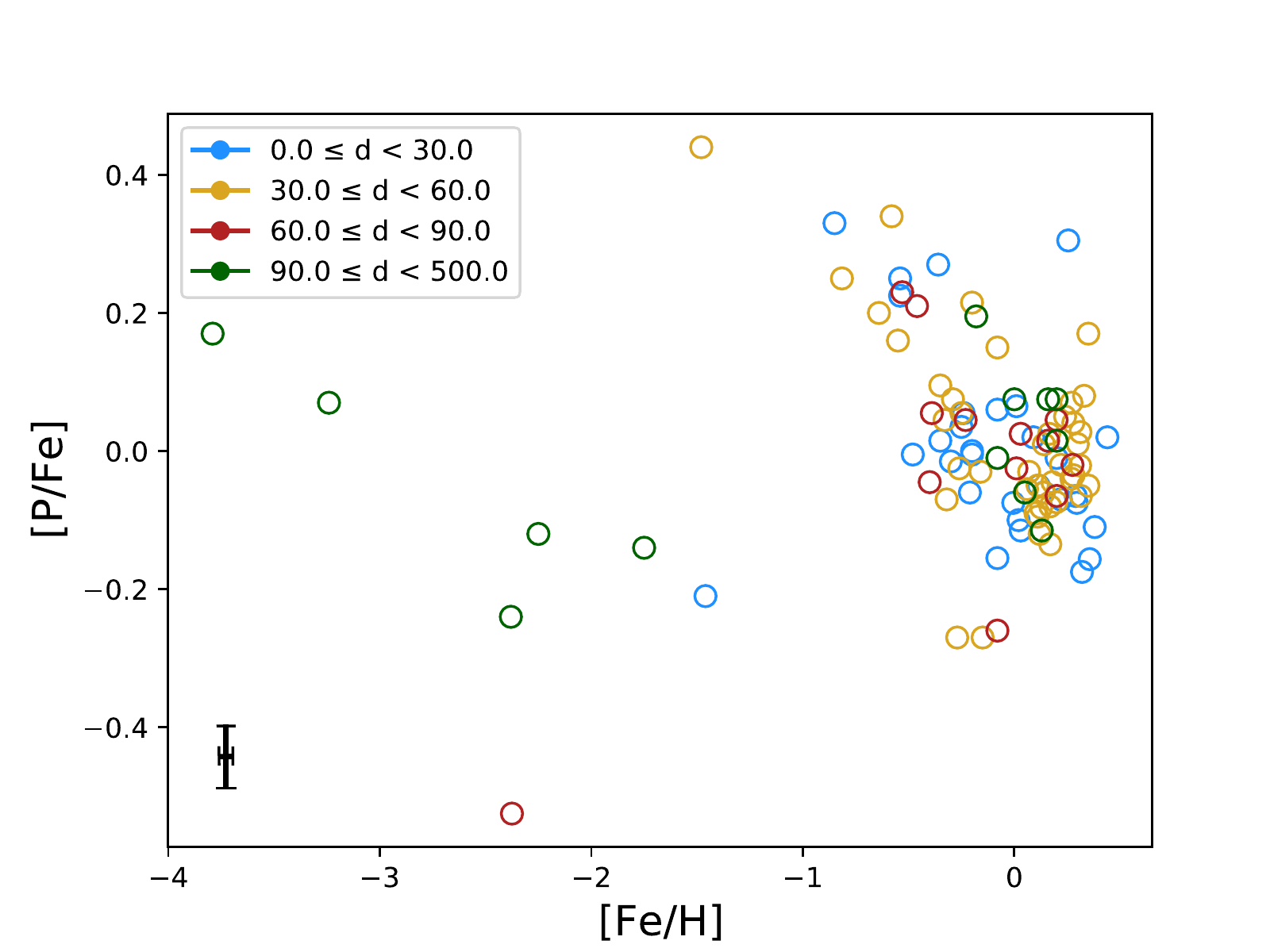}
    \caption{The 100 stars within the Hypatia Catalog that have phosphorus abundance measurements, with respective error bar in the bottom left corner where typical [Fe/H] error is 0.04 dex while [P/Fe] is 0.11 dex}. Stellar elemental abundances are reported in dex notation, i.e., a logarithmic ratio of elements in the star compared to the same ratio in the Sun. The stars are color-coded to indicate distance.
    \label{fig:PvsFe}
\end{figure}

Only seven groups have successfully measured phosphorus stellar abundances, since the absorption lines fall outside of the optical band typically used for spectroscopy. As a direct result, there are a total of 100 phosphorus abundances within FGKM-type stars -- or $\sim$1\% of stars within Hypatia (Fig. \ref{fig:PvsFe}). Infrared abundances, i.e., from the 1053.24 and 1058.44 nm lines, were determined by \citet{Caffau11, Caffau15b, Caffau19}, with 0.04 dex, 0.04 dex, and 0.12 dex  average uncertainties, respectively. Similarly, although using the 1058.1 and 1059.6 nm lines, \citet{Maas17} measured phosphorus in 22 stars, with an average 0.07 dex uncertainty and \citet{Maas19} had an average uncertainty of 0.08 dex in 21 stars from the disk and Hyades cluster. \citet{Masseron20} used the APOGEE survey DR14 to measure neutral P lines at 1571.15 and 1648.29 nm for 30 stars likely originating from the Galactic thick disk or halo, with an average uncertainty of 0.15 dex. \citet{Roederer14b} and \citet{Jacobson14J} looked in the near-ultraviolet between 0.2135-0.2555 nm utilizing a variety of neutral P lines that resulted in abundances with an average 0.29 dex uncertainty. Figure \ref{fig:PvsFe} shows all of the [P/Fe] measurements with respect to [Fe/H] in the Hypatia Catalog, where the data is color-coded to indicate distance. The majority of the data is centered around the 0.0 dex value, which is the same as the Sun. There is significant scatter in the data. A weak trend of decreasing [P/Fe] vs. [Fe/H] in thin disk stars has  been identified previously \citep{2019AJ....158..219M}. This is consistent with co-production of P and $\alpha$ elements (see Section~\ref{sec:future} for further discussion of nucleosynthetic sources of P).

\section{Stellar Abundances \& Molar Ratios}
Of the 100 stars within the Hypatia Catalog with P abundance measurements, the majority are G-type stars (64 stars with 5000 $\le$ \teff $<$ 6000 K), with notably fewer F-type stars (33 stars, 6000 $\le$ \teff $<$ 7500 K) and K-type stars (3 stars, 3500 $\le$ \teff $<$ 5000 K). There have not been any P abundance measurements M-type stars. In comparison, there are 6072 stars in the Hypatia Catalog with C abundances, 2927 with N abundances, and 7878 stars with Si abundances -- not including the instances when multiple groups measured the same element within an individual star. We are particularly interested in P, N, and C because they are primary elements required for life on Earth. As the limiting nutrients for central (also known as core) metabolism, the presence of P and N in appropriate proportions is potentially critical for determining the habitability of exoplanets.  We also included ratios relative to silicon, in part because Si is biologically relevant, but also because Si is a major rock-forming element on Earth and is more frequently measured in stars. 

\begin{figure}
    \centering
    \includegraphics[width=1.0\linewidth]{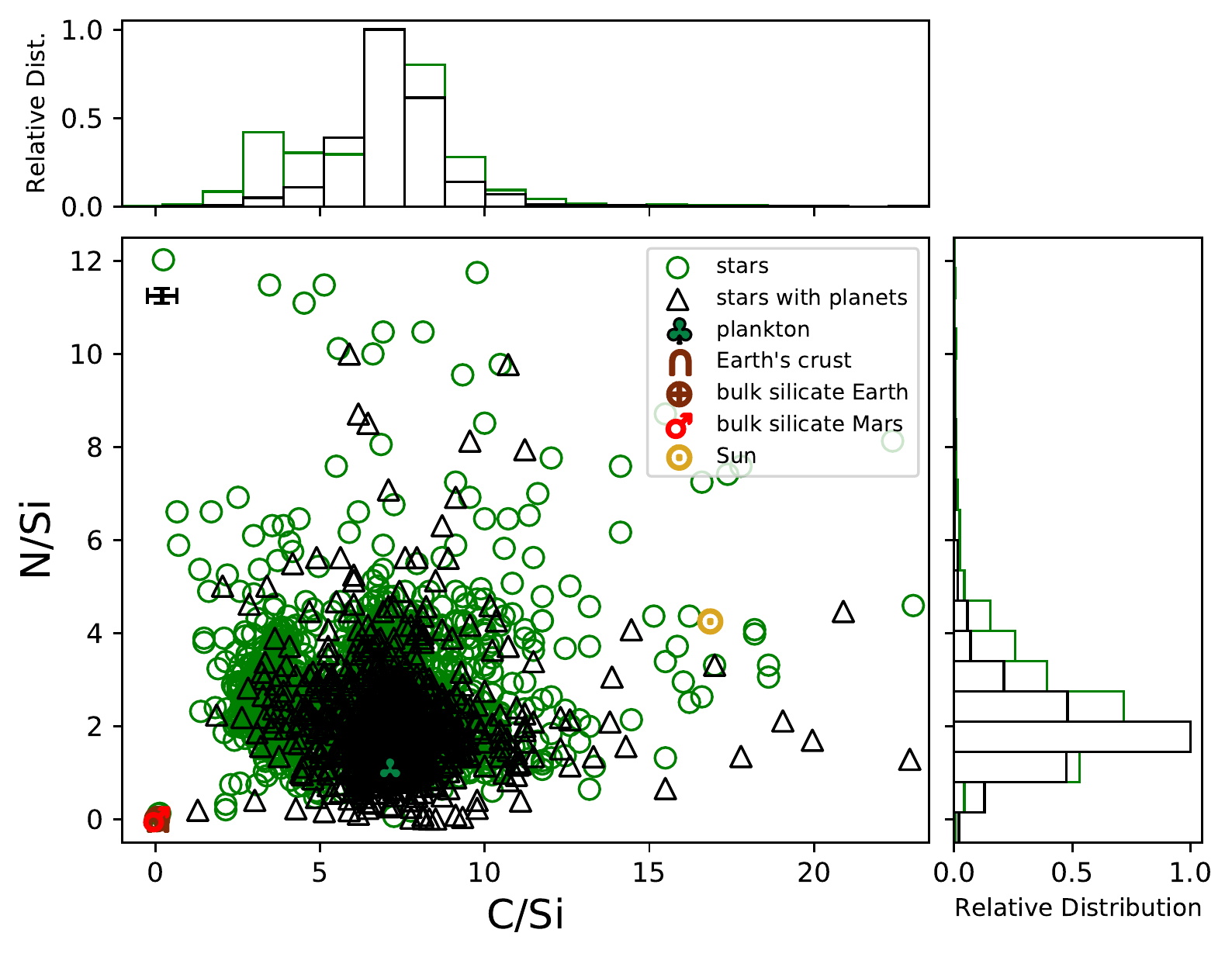}
    \caption{Molar ratios, C/Si versus N/Si, for stars within Hypatia that have C, N, and Si abundance measurements. There are 2818 stars (green circles) of which, 1008 are exoplanet host stars (black triangles) -- note that 22 data points lie beyond the plot boundaries to better focus on the majority of the data. The green leaf, brown Earth symbols, red Mars symbol, and yellow Sun symbol indicate the elemental ratios for these objects, respectively. The histograms summarize the relative distribution of elemental ratios in the Hypatia stars (green) and exoplanet hosts (black). The representative error bar is located in the top left corner.}
    \label{fig:green}
\end{figure}

\begin{figure*}
    \centering
    \includegraphics[width=0.45\linewidth]{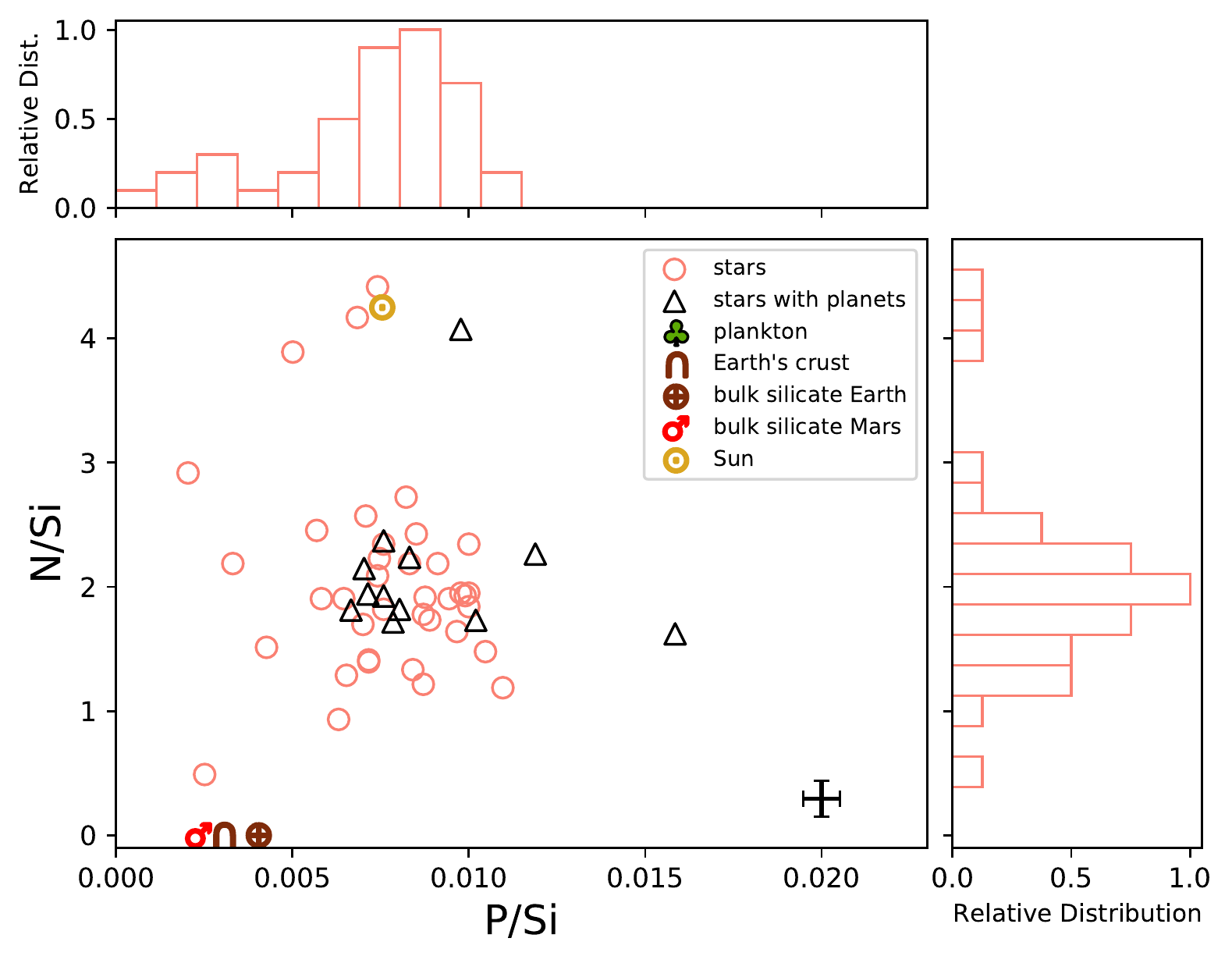}
    \includegraphics[width=0.48\linewidth]{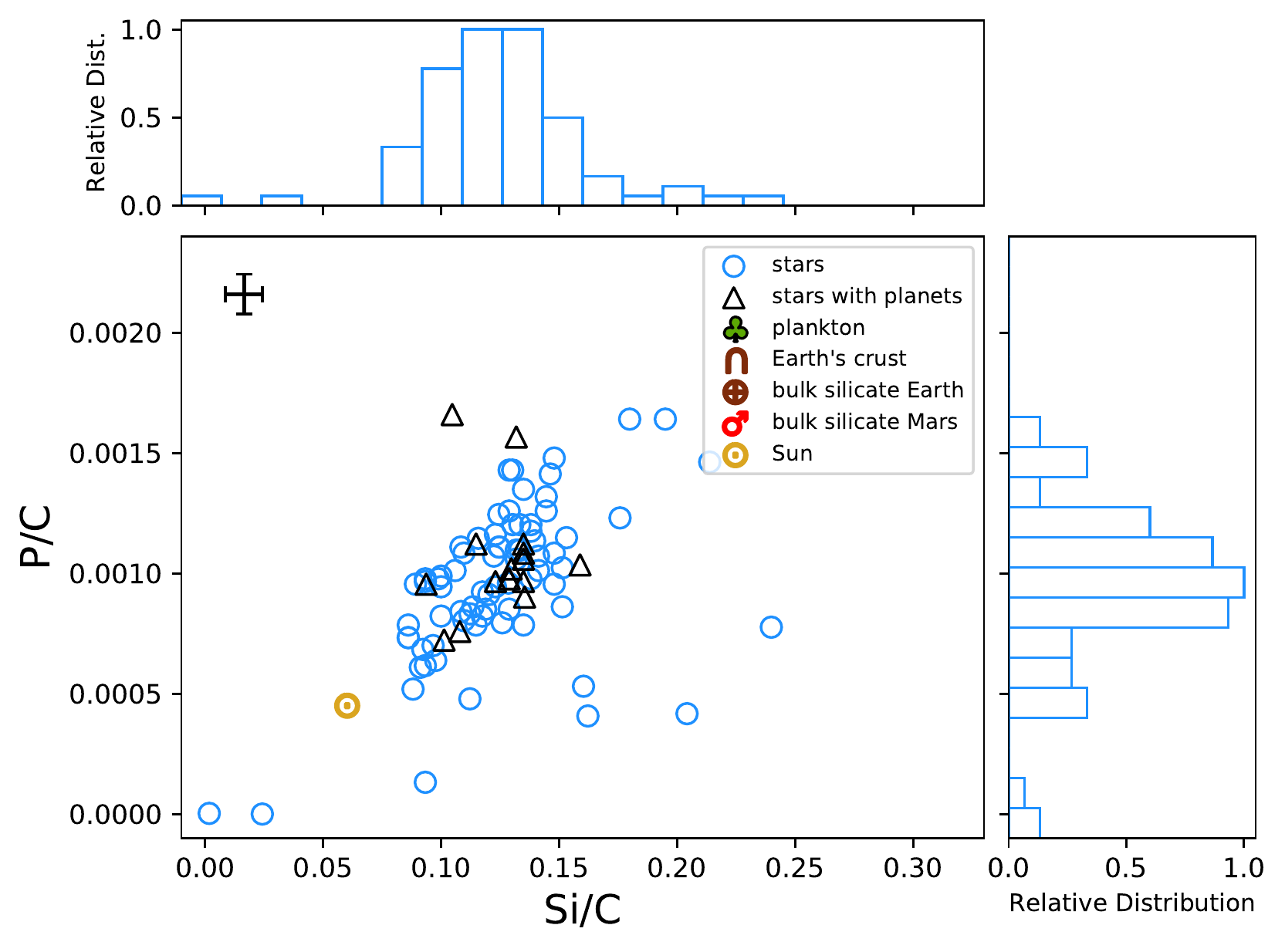}
    \caption{Similar to Fig. \ref{fig:green}, but the left plot, in salmon, shows the molar ratios of P/Si versus N/Si; a representative error bar is shown in the bottom left corner. Note, two stellar data points have been omitted at high N/Si as well as the relative distributions of the exoplanet host stars given the smaller numbers. Also, the plankton data point is off-scale to the right at P/Si = 0.07 and N/Si = 1.07. The right plot, in blue, shows Si/C versus P/C; a representative error bar is shown in the top left corner. The points for Earth, Mars, and plankton all plot at much higher Si/C and P/C values than the stars.  }
    \label{fig:blue}
\end{figure*}

\begin{figure*}[t!]
    \centering
    \includegraphics[width=1.0\linewidth]{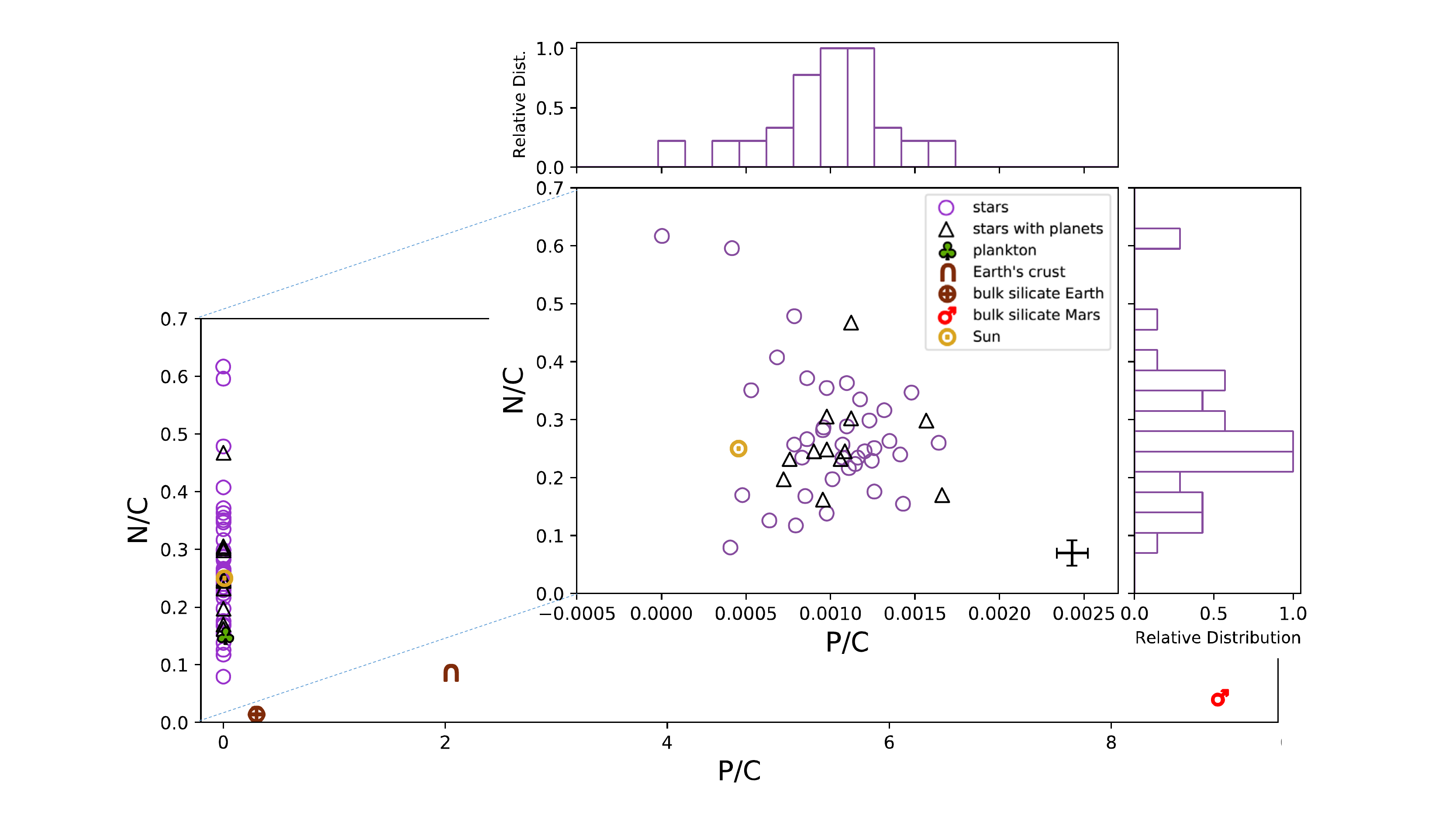}
    \caption{Similar to Fig. \ref{fig:blue}, but with molar ratios for  P/C and N/C; a representative error bar is located in the bottom right corner of the inset figure. We have displayed the data using two x-axes, in order to show the range in Earth and Mars comparators as well as the scatter in the stars.} 
    \label{fig:purple}
\end{figure*}

In Figures \ref{fig:green}-\ref{fig:purple} we show the correlations between C, N, P, and Si. We have opted not to show these elements in dex notation, but instead as molar ratios, A/B. These are not mass ratios but chemical ratios that provide information about the relative stoichiometry of objects (i.e., an empirical formula) and allow us to compare them to that of a specific chemical reaction, in this case photosynthesis.
Not only are molar ratios more commonly used by Earth scientists, i.e., biologists, chemists, and geologists, but this notation also indicates how much material is available to react -- an important and dominant process, even in stars (e.g., the CNO process). Therefore, in an effort to bridge interdisciplinary fields, we have removed the solar normalization in lieu of a notation that is more meaningful when considering the chemistry of planets. 
Mathematically, to convert from dex notation to molar ratio: 
\begin{equation}\label{dex2moles}
X/Y = 10^{( \mathrm{log}\, \epsilon(X)_{\odot}+[X/H]_*)} /\, 10^{( \mathrm{log}\, \epsilon(Y)_{\odot}+[Y/H]_*)}, 
\end{equation}
\noindent
where $\odot$ indicates the solar normalization for that element and $*$ designates the abundance measurements in dex relative to the Sun. While there are dozens of solar abundance scales \citep{Hinkel14}, Eq. \ref{dex2moles} removes the solar baseline so there is no longer any dependence on the adopted solar abundance. Because all stellar abundances in the Hypatia Catalog are normalized with respect to \citet{Lodders09}, we use their absolute solar abundances when converting from dex notation to molar ratios. 

The molar ratios of N/Si as a function of C/Si are shown in Fig. \ref{fig:green}. There are 2818 stars, 1008 of which host exoplanets, in Hypatia that have determinations for all three elemental abundances.
As indicated by the histograms on the x- and y-axes, the majority of the stars have C/Si molar ratios within 5-10 and N/Si $\approx$ 1-3. In comparison, the majority of exoplanet host stars (black) span a smaller range of N/Si. Looking at the distribution of planets in our sample, we find that the planet population is fairly random -- varying from smaller Earth-sized planets to mini-Neptunes and Jupiter-sized planets. For example, the median planet radius is 0.19 R$_J \approx$ 2.1 R$_{\oplus}$, with a minimum at 0.03 R$_J \approx$ 0.33 R$_{\oplus}$ and maximum at 2.0 R$_J \approx$ 22 R$_{\oplus}$. The majority of planets have a radius between 0.05 R$_J \approx$ 0.56 R$_{\oplus}$ and 0.03 R$_J \approx$ 3.4 R$_{\oplus}$. The median planet mass is 0.75 M$_J \approx$ 238 M$_{\oplus}$ with a minimum of 0.002 M$_J \approx$ 0.64 M$_{\oplus}$ and a maximum of 33 M$_J$. However, most of the planets have masses that fall between 0.01 M$_J \approx$ 3.2 M$_{\oplus}$ and 0.04 M$_J \approx$ 12.7 M$_{\oplus}$.

Due to the low number of P determinations, the overlap of stars with multiple bioessential measurements is quite low. Fig. \ref{fig:blue} (left) relates the molar ratios P/Si and N/Si for 54 stars, including 12 exoplanet host stars. There are no obvious trends in either population, although most stars have P/Si between 0.007-0.011 while the majority have N/Si between 1-2.5. 
There are 93 stars for which C, P, and Si have all been measured (Fig. \ref{fig:blue}, right), where 16 of those stars host planets. Despite a smaller number of stars than in Fig. \ref{fig:green}, the general trend for the majority of stars is that increasing Si/C is accompanied by a proportionate change in P/C. This is likely indicative of different production sites for the elements, since P is primarily produced in massive stars whereas C is created in AGB stars and Si is made when massive stars become Type Ia supernovae. Since P and the majority of Si are produced in massive stars, their abundances are more strongly correlated with each other than with C.  

Similar to Fig. \ref{fig:blue} (left), there are only 54 stars, including 12 that host planets, with C, N, and P abundances in Fig. \ref{fig:purple}. Again, there is no obvious difference between the C/N and C/P ratios of stars that host planets and those that do not, except perhaps that planet-hosting stars tend to have C/P $<$ 1500 -- but this may be due to the relatively small number of measurements. Of the planet hosting stars, 7 have been confirmed to have multiple planets. Four of the systems have wide-orbiting planets with periods longer than 1000 days. The median planet mass is 0.73 M$_J \approx$ 232 M$_{\oplus}$; the majority of planets have masses between 0.0076 M$_J \approx$ 2.4 M$_{\oplus}$ and 0.06 M$_J \approx$ 19 M$_{\oplus}$ or from roughly super-Earth to Neptune-sized planets.

The scatter in Figs \ref{fig:green}--\ref{fig:purple} derives from the measurements themselves; i.e., from a combination of observational error and intrinsic physical variation. The observational error can be statistically removed to determine the intrinsic variation in elemental ratios of the sample \citep{2014AsBio..14..603Y, 2018haex.bookE..60Y}. 
Within a given survey, the method is as robust as the reported uncertainties. It evaluates each data point with individual errors (if reported) and is designed to deal with heteroscedastic (varying uncertainty) data. Systematic shifts between surveys can be dealt with by evaluating the intrinsic spread in each survey independently. However, P is a difficult case because of the small numbers in total as well as within any given survey. As there are no major systematic shifts apparent by inspection, we compared the intrinsic spread produced by analyzing the entire catalog as a single set with the larger survey samples. They are consistent, but we have avoided quoting a precise value because of this consideration. When observational error is removed, C, N, and P all display real and substantial variations in their abundance ratios in stars, as the observational error is much smaller than the measured spread.

\begin{table*}
\caption{Molar Ratios Normalized to P}
\begin{center}
\begin{tabular}{|l|cccc|c|}
\hline
\hline
 & C & N & Si & P & Reference \\
\hline 
Plankton & 106.0 & 16.0 & 15.0 & 1 &  \citet{Redfield1958} \\ 
Earth Crust & 0.49 & 0.04 & 291.16 & 1 &  \citet{masonandmoore82} \\ 
Bulk Silicate Earth & 3.44 & 0.05 & 2573.35 & 1 &  \cite{mcdonough03} \\ 
Bulk Silicate Mars & 0.11 & 0.005 & 443.89 & 1 &  \cite{yoshizaki20} \\ 
Sun & 2233.93 & 564.82 & 132.88 & 1 & \citet{Lodders09} \\ 
Hypatia Catalog Star (average) & 3814.33 & 1010.77 & 235.79 & 1 &  \citet{Hinkel14} \\
\hline
\hline
\end{tabular}\label{tab.molarratios}
\end{center}
\end{table*}

Each of Figures \ref{fig:green} through \ref{fig:purple} include molar ratio values for marine plankton (green $\clubsuit$), Earth's crust (brown $\cap$), bulk silicate Earth (brown $\oplus$), bulk silicate Mars (red $\mars$), and the Sun (yellow $\odot$). These values and their references are summarized in Table \ref{tab.molarratios}, and are all shown on the same scale in Fig. \ref{fig:purple}. We chose to present Earth's crustal ratios because the crust is ultimately the only source of P to organisms. The Earth's crust has a  composition distinctly different from that of the bulk silicate Earth (BSE; e.g., the mantle and crust) because crustal differentiation imparts a strong chemical fractionation. In addition, the mantle is a much larger fraction of the mass of the planet compared to the relatively thin crust, so the BSE is dominated by the mantle composition. We also compare different planets using calculated ratios for the bulk silicate Earth and bulk silicate Mars. We did not include Earth's atmosphere in this analysis because, despite being a significant pool of nitrogen, the atmosphere contains a very small fraction of the Earth's C and N and there is no significant P in the atmosphere. We also did not include Earth's core in the analysis because the data is less well constrained; however, this could be an interesting line of future research. 

Table \ref{tab.molarratios} also gives C, N, P, and Si values for all stars in the Hypatia Catalog where individual measurements for each element (from 6072, 2927, 100, and 7878 stars, respectively) were averaged. 
Note the sample of stars in Figs. \ref{fig:green}--\ref{fig:purple} is slightly different, since the table required stars with measurements for all four elements shown. 
The solar abundances in Table \ref{tab.molarratios}, were taken from \citet{Lodders09};  the Sun's molar ratios are, notably, end-members of the distributions in Figs \ref{fig:green}--\ref{fig:purple}. However, when analyzing the Sun with respect to similar ``twin'' stars that have similar effective temperatures, surface gravities, and [Fe/H] abundances, this is not particularly surprising. \citet{Bedell18} performed a careful analysis of the abundances within solar twins and found that, in comparison, the Sun is relatively deficient compared to 96\% of the sample in its refractory-to-volatile ratio. In other words, the Sun either has lower abundances of refractory elements, roughly equivalent to 4 Earth masses of rocky or chrondritic material \citep{Chambers10} or it has a surplus of volatile elements (e.g., C and O). 

\section{Biological \& Geological Implications}
We have considered several components of the Earth  system, namely, the bulk silicate Earth -- i.e., the composition of the primitive mantle after accretion and the separation of siderophile elements to the Earth's core, Earth's crust -- i.e., the region that interacts  with biology, and plankton -- a proxy for photosynthetic life on the Earth's surface, because the differentiation of planets and the evolution of life each impart a chemical fractionation. In addition, we have included bulk silicate Mars to illustrate compositional differences for planets in the same system. In Figs \ref{fig:green} and \ref{fig:blue} (right), the Earth and Mars comparators have molar ratios that are deficient with respect to the Sun and the other Hypatia stars. In Fig. \ref{fig:blue} (left), the P/Si ratio for bulk silicate Earth and Mars, Earth's crust, and the Sun were all similarly close to P/Si = 0.0, while the Hypatia stars have a range P/Si $\approx$ 0.005-0.01 and plankton has P/Si = 0.67. Notably, also, the Sun has N/Si = 4.25, while plankton has N/Si = 1.07 and the Earth and Mars components are very close to N/Si = 0.0. In contrast, Fig. \ref{fig:purple} shows that the Earth and Mars comparators have a larger range in C/N compared to the Sun and the Hypatia stars. However, bulk silicate Earth and Mars and Earth's crust have a lower N/P ratio than plankton, the Sun, or Hypatia stars (see Table \ref{tab.molarratios}). Overall, the Sun appears to have a C/N ratio similar to the Hypatia average; however, the Sun has C/P and N/P ratios that are a factor of two lower than the Hypatia average, strongly suggesting that our Sun is relatively P-rich. While individual stars may have larger error bars, there is no reason to believe that the entire sample is systematically shifted, such that the position of the main locus of stars relative to the Sun should be sufficiently accurate to confidently identify the Sun as unusual. The Earth and Mars ratios indicate they are N-poor and P-rich, relative to the Sun and to the Hypatia stars. One hypothesis is that N, a volatile element, is preferentially segregated towards gas giant planets while P, a strongly siderophile element, is thought to be condensed into the cores of rocky planets \citep{Stewart07}.

Life's stoichiometry is not perfectly fixed either. The C/N/P ratio in plankton biomass can vary, somewhat, depending on the species and the supply of elements to the particular environment. For example, bacteria are more plastic in their elemental ratios and can have C/N ratios as low as 3 or 4. Trees, in contrast, are N-poor relative to carbon and can have C/N ratios as high as 20 or so. However, the range in C/N ratios is not especially large. Phosphorus is generally a more constant fraction of biomass, presumably due to the absolute need for P in genetic material.

While elemental abundance ratios are an important consideration for astrobiology, molar ratios dictate the potential for chemical reactions to occur. We should note that speciation, the actual chemical form of the element, is also significant. If N is only present as N$_2$ gas, but not as nitrate or ammonium, the planet might be less likely to host life; though we note on Earth some microbes can fix N$_2$ from the atmosphere. The speciation of P also matters. On Earth, orthophosphate, \phos, is the bioavailable form of P and the predominant form in crustal minerals. If reduced P is present, for example as Fe$_3$P which may be prevalent in Earth's core, it may not be useful from a biological perspective. However, even within a single planetary system, much like the Solar System, we would expect planets that formed within different parts of the protoplanetary disk would range in core size, mantle composition, and bulk density. Yet, with such a dearth of information for P, it is difficult to make any assessment of P's role within an individual exoplanet's geologic evolution or geochemical cycles, and impossible when it comes to making an assessment of rocky exoplanets in general. However, even knowing the abundance ratios is a great step forward for assessing whether stars might supply appropriate amounts of bioessential elements.


\section{Future Considerations \label{sec:future}}
Phosphorus is the most critical element for life and it is three orders of magnitude less abundant in the Sun than other important light elements. In addition, measuring P in other stars is extremely difficult, since there are no P absorption lines in the optical band. Even lines in the near-ultraviolet are often blended and are difficult to discern from the stellar continuum \citep{Roederer14b}; lines in the infrared are also fairly weak and are blocked by atmospheric tellurics \citep{2017AJ....154...94M}. There have been a handful of successful P abundance analyses. However, these studies were specifically targeted at P and required not only extremely hard work, but also had to contend with a limited number of suitable spectrographs available, particularly for the near-IR lines. Therefore, to improve our understanding of P and its role in planetary habitability, the community must develop instrumentation that can overcome these observational challenges. For example, an infrared spectrograph in low-Earth orbit or in space would be able to access molecular P lines (such as PS) as well as elemental lines that aren't attainable from the ground.

In the absence of direct P measurements, it may be possible to obtain rough constraints on P ratios from proxy species. The odd-Z elements Al, P, K, and Sc are all produced primarily in massive stars, though Al has a small component from intermediate mass stars. Phosphorus is produced primarily in hydrostatic O burning, ${\rm ^{16}O(^{16}O,p)^{31}P}$, and free neutron captures in O and Si burning. Aluminum is produced during O burning and C/Ne burning; similarly, K and Sc are produced by free neutron captures primarily in Si burning \citep[e.g.][]{1996snih.book.....A}. This partial co-production of Al, P, K, and Sc makes these candidates for proxies. The ratios of P to these elements, P/X, to the ratios of the proxies to Fe, (Al, P, K, and Sc)/Fe, is relatively tightly correlated. Based on the sparse observational data available, the ratios P/K versus K/Fe and P/Sc versus Sc/Fe have a scatter of order 60\% at [Fe/H] $> -1$ arising from observational error and stellar intrinsic variation. Normalizing the ratios of P/Al to Mg/Fe allows the contribution of Type Ia supernovae to be removed \citep{Timmes:1995p3197, 2019AJ....158..219M}, flattening any trend lines but does not significantly reduce the scatter, such that using P/Al gives a scatter of a factor of two. Of these species, Al is the most easily measured, but Sc is also frequently reported. Measurements of K in the literature are significantly rarer, but more common than P by a factor of about ten \citep{Hinkel14}.

It is clear from the literature that it is difficult to measure P within stars, to the extent that it's not possible to outline a plausible range or generalization of P abundances within nearby stars, let alone exoplanet host stars. Even N abundance measurements, which are more common, have only been achieved in $\sim$31\% of Hypatia stars. From Figs \ref{fig:green}--\ref{fig:purple}, it would seem that other stars are enriched in C and N with respect to the Sun. However, the Sun may not be a typical star in terms of its composition, per Table \ref{tab.molarratios}. While the Sun is the only star around which there is confirmed life, its outlier status among other stellar `twins' makes it an odd choice on which to normalize elemental abundances, [A/B]. It could be argued that the proximity of the Sun allows us to measure its composition to a higher accuracy, however one of the functions of the Hypatia Catalog is to renormalize abundances to the same solar scale. As a result, there are +60 solar normalizations that have been collected to-date, where the range in solar Fe is $\Delta$ $\log \epsilon$(Fe) = 0.26 dex. This is regardless of typical uncertainty for [Fe/H] = $\pm$0.05 dex and the fact that Fe is usually considered so well known that it is often used by astronomers as a proxy for the abundance of all heavy elements, or overall ``metallicity," in a star. Therefore, when carefully defining a chemical range that is likely to beget a habitable planet, it may make the most sense to use another star as reference or even no reference at all. By varying the standard way in which we present stellar abundance data, we may begin to remove some Solar System-centric assumptions while also bridging the gap between different disciplines important for understanding exoplanetary habitability.

To date, current exoplanet data is biased towards large, gaseous planets that are easier to detect around main-sequence stars. However, with the continuation of TESS and the launch of JWST, the Roman Telescope, and ARIEL, the gaps in known exoplanet demographics will begin to close. In anticipation of this new data, it is essential that planet formation models are able to reconcile how elements go from the stellar host to the planet and then differentiate, i.e., between the core and the surface, once they are in the planet. We know that specific molar ratios of C/O and Mg/Si are more conducive to habitable planets than others, such that C/O ratios $\sim$0.8--1.0 are likely to produce geodynamically inactive planets \citep{Unterborn14, bond_2010_aa}. However, P is required for life, thereby placing a lower bound on what is necessary in the star to go from star to planet to life. For example, if there is very little P in the stellar birth cloud, then there won't be much to add as a planetary veneer. Or what little P is available during planetary formation could all go into cores, with only an insignificant amount able to escape to the planetary surface. Fortunately, on Earth, biology has evolved to function with very little available P. However, if there are stars with practically insignificant amounts of P, then their planets are likely inhospitable for life; perhaps to the extent that we could rule out the possibility of life altogether on the planet's surface. In the truest sense, it is absolutely vital to understand planetary bulk composition, internal structure, mineralogy, and atmosphere \citep[e.g.,][]{Foley18} in order to fully assess whether a planet is habitable. 

In this paper, we have provided an example where an understanding from geobiology reveals that P, an element thus far underappreciated within astrophysics, is critically important for biology. On the other hand, the geobiologist learns that P is incredibly difficult to measure, especially in context with other bioessential elements. Both parts of the interdisciplinary collaboration will need to work together to advance the thinking of the exoplanet community and get the data needed to trace biological systems on exoplanets. They will need to overcome jargon and assumptions, for example the treatment of H within the fields, the role of major/minor/trace elements, the definition of stellar ``metallicity" as [Fe/H] or as the summation of the heavy elements \citep{Hinkel19}. All of these concepts are critical to developing more interdisciplinary, exoplanet-based science such that we can define planetary habitability from a holistic perspective that includes astronomy, biology, geology, and chemistry.

\acknowledgments 
The results reported herein benefited from collaborations and/or information exchange within NASA's Nexus for Exoplanet System Science (NExSS) research coordination network sponsored by NASA's Science Mission Directorate. HEH acknowledges NASA support from grant \#NNX15AD53G. The research shown here acknowledges use of the Hypatia Catalog Database, an online compilation of stellar abundance data as described in \citet{Hinkel14} that was supported by NASA's Nexus for Exoplanet System Science (NExSS) research coordination network and the Vanderbilt Initiative in Data-Intensive Astrophysics (VIDA).

\end{document}